\documentclass[12pt]{article}

\usepackage{sbc-template}
\usepackage{graphicx,url}
\usepackage{listings}
\usepackage{xcolor}
\usepackage[utf8]{inputenc}

\newcommand{\code}[1]{\texttt{#1}}

\title{Finding new routes for integrating\\Multi-Agent Systems using Apache Camel\thanks{Supported by Petrobras project AG-BR, IFSC and UFSC.}}

\author{Cleber J. Amaral\inst{1,2}, S{\'e}rgio P. Bernardes\inst{1}, Mateus Conceição\inst{1}\\Jomi F. Hübner\inst{1}, Luis P. A. Lampert\inst{1}, Ot{\'a}vio A. Matoso\inst{1}, Maicon R. Zatelli\inst{1}\\}

\address{Universidade Federal de Santa Catarina (UFSC)\\
  Florian{\'o}polis -- SC -- Brazil
\nextinstitute
  Instituto Federal de Santa Catarina (IFSC)\\
  S{\~a}o Jos{\'e} -- SC -- Brazil\\
  \email{\vbox{\noindent cleber.amaral@ifsc.edu.br, \{sergiopb1998,lp.lampert\}@gmail.com
    mateusconceicao1@hotmail.com, \{jomi.hubner,maicon.zatelli\}@ufsc.br
    otaviomatoso@yahoo.com.br}}
}

\begin{document} 

\maketitle

\begin{abstract}
In Multi-Agent Systems (MAS) there are two main models of interaction: among agents, and between agents and the environment. Although there are studies considering these models, there is no practical tool to afford the interaction with external entities with both models. This paper presents a proposal for such a tool based on the Apache Camel framework by designing two new components, namely \emph{camel-jason} and \emph{camel-artifact}. By means of these components, an external entity is modelled according to its nature, i.e., whether it is autonomous or non-autonomous, interacting with the MAS respectively as an agent or an artifact. It models coherently external entities whereas Camel provides interoperability with several communication protocols.

\end{abstract}

\section{Introduction}

MAS literature has plenty of research about agents' interactions, i.e., agents sending and receiving messages to and from other agents (A-A). Many approaches model almost any entity as an agent and thus the interaction remains something among agents. However there are new approaches that questioned the \emph{agentification} method proposing an MAS where non-autonomous entities are conceived as artifacts in the environment. In these approaches, the development of an MAS considers the design of both agents and artifacts. The environment is not simply what is outside the system (the exogenous environment), but it is designed accordingly to the system purpose (the endogenous environment)~\cite{Ricci2006, omicini:jaamas08}. 
In this sense, we have two models of interactions: agent-to-agent (A-A) and agent-to-environment (A-E). In the former, an agent \emph{communicates} with another agent using an Agent Communication Language (ACL) and in the latter, an agent \emph{perceives and acts} upon artifacts in the environment.

When we consider the integration with other applications, those two models are adopted by the current development platforms. On the one hand, we have approaches that use ACL for that purpose and other applications are seen by the agents as other agents (having a mental state, implied by the ACL semantics). On the other hand, we have approaches where other applications are seen as part of the environment and agents perceive and act on them. Some platforms provide an A-A approach while others an A-E approach, but, as far as we know, no platform provides both. The designer is forced to conceive some other application either as an agent or as an artifact, despite the application properties. 

In this paper, we propose to apply the same argument as~\cite{omicini:jaamas08} for the integration of MAS and external applications: some external applications are autonomous and should be modelled as agents while others are non-autonomous and should be modelled as artifacts. 

For instance, in the Industry 4.0 context, it is expected the interaction of many entities such as an autonomous planner sending commands to a non-autonomous machine, which signalises what was done. Later, the planner must choose a supplier after an auction to hire a freight to take the product to the destiny, which is usually an human. This short example gives an idea of how comprehensive and challenging the integration can be. We can notice that both models of integration are required: the \emph{autonomous} planner and the human are better modelled as agents, performing A-A interactions, and the \emph{non-autonomous} machine should be integrated as an artifact which when communicating with and agent performs and A-E interaction. Following this concept, we have developed two components for JaCaMo platform, for integration among agents, and between agents and the environment. The referred components are used to set communication routes for the MAS and external entities, using the framework Apache Camel~\cite{Ibsen:2010:CA:1965487}, a mediation tool to provide interoperability with many technologies.

\section{MAS integration approaches}

MAS are being applied as a core technology for distributed systems that needs cooperation and negotiation~\cite{Roloff2016}. The integration of MAS and external entities, i.e. any entity which was not defined its totality within the MAS itself, regards concerns such as compatibility with standards, interoperability and portability. We have found two main forms of integration: (i) among agents (A-A); and, (ii) between agents and the environment (A-E).

\subsection{Integration among agents (A-A)}

The communication among agents is usually done by speech-acts which considers utterances as actions, usually intending to change the mental state of recipients. The utterance can inform beliefs, desires and intentions of rational agents that attempt to influence other agents. The Knowledge Query and Manipulation Language (KQML) is the first speech-act based language providing high level communication in the distributed artificial intelligence applications~\cite{Vieira2007}. In fact, once speech acts became widely accepted in MAS community, the integration among different agent's platforms was facilitated.

Currently FIPA-ACL, which is very similar to KQML, is the main standard for agents communication. FIPA-ACL uses performatives to make explicit an agent’s intention for each sent message, for instance, \emph{inform} is used to influence the recipient to believe in something, and \emph{request} to influence the recipient to add something as a goal~\cite{Vieira2007}.

Another communication aspect is related to the expected sequence of messages. Conversations among agents usually follow some patterns which are often referred to as interaction protocols. Typical patterns such as negotiation, auction, and task delegation are defined using FIPA standards~\cite{Bellifemine2005}. In addiction, there are communication infrastructures that allow agents to be distributed over a network. The challenge in A-A is the integration between an agent, for instance, using FIPA-ACL, and another agent using another language, for instance, an human. This situation leads to the necessity of some tool to make both end-points compatible.

\subsection{Integration between agents and the environment (A-E)}

There are systems or parts of a system which are better seen as resources or tools that can be used by agents to achieve their goals. These entities, called artifacts, have no internal goals, they are not autonomous and neither proactive, but they supply useful functionalities for agents. An analogy for agents and artifacts is the interaction between humans, as autonomous entities, and tools they exploit in their activities~\cite{Ricci2006}. For instance, a blackboard shared by agents would be modelled as an artifact, being predictable and deterministic, if not, it would perform undesirable autonomous behaviour.

Artifacts are placed in workspaces which represent areas of the MAS environment. Agents can perceive changes and act within the workspaces they are occupying, and on artifacts they are watching, i.e., being aware of events and signals due to interactions with non-autonomous entities held inside of virtual boundaries. The environment can reflect the effects of agents' actions and other phenomena. It is being treated as a first-class programming abstraction with similar importance of agents programming abstraction~\cite{Ricci2006}.

Besides modelling non-autonomous entities, the artifacts can also be used for other purposes: (i) for agents coordination using shared artifacts such as organisational boards or coordination marks; (ii) for indirect communication among agents, for instance, by blackboard artifacts; (iii) for implementing the user interface of a system; (iv) for controlling transactions over environment elements through distribution and synchronisation facilities; and (v) for integration between the MAS and external entities~\cite{boissier_bordini_hubner_ricci_2019}.

Regarding the use of artifacts to integrate external entities, the integration is done usually through specific Application Programming Interfaces (APIs). The main concern with this approach regards to the high programming effort when there are different protocols in scenarios of heterogeneous devices.

\section{Integrating A-A and A-E using Camel}

\begin{figure}[ht]
\centering
\includegraphics[width=0.8\textwidth]{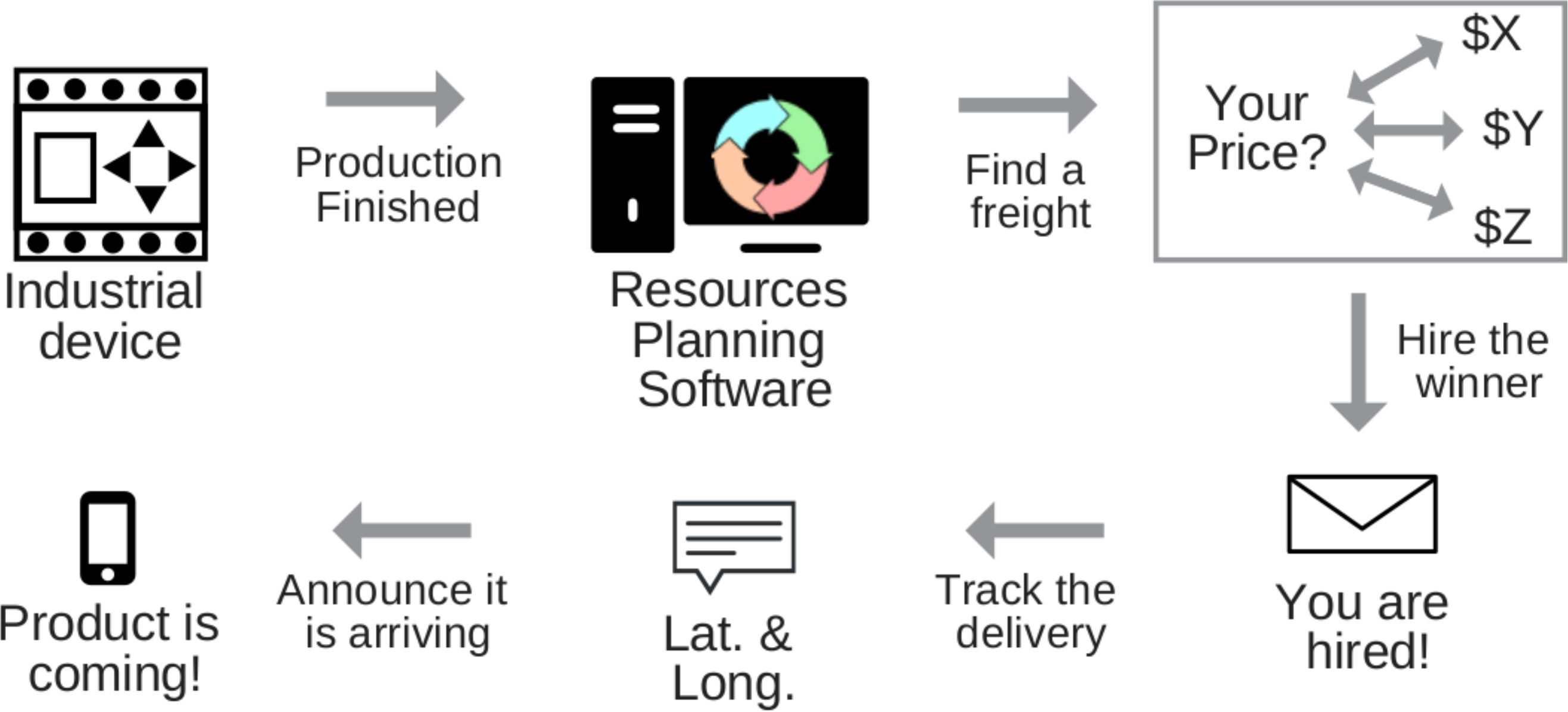}
\caption{Finishing production and delivering the product in Industry 4.0 context}
\label{fig:exampleFig1}
\end{figure}

Back to our example in the Industry 4.0 context, Figure~\ref{fig:exampleFig1} shows a process that begins with a packed product, in a production line, up to its delivery to the customer. On the first step, there is an industrial device, a non-autonomous entity, that communicates using an industrial protocol. Once the device signalises the end of the production, the order is checked out on the Enterprise Resource Planning (ERP) software, another non-autonomous entity. The supplier should choose the best offer for a freight, which may be done by accessing suppliers' systems to then interact with the winner, an autonomous entity. Later, the delivery should be tracked by a monitoring system, which is non-autonomous. Finally, when the product is near the destination, a message must keep the client, an autonomous entity, informed.

This scenario illustrates the requirements for the integration: heterogeneous endpoints with different kinds of interactions due to the autonomous and non-autonomous nature of entities. We propose the use of the framework Apache Camel for both integration models, A-A and A-E. We have thus two components: \emph{camel-jason} for integrating MAS's internal agents with external entities modeled as agents, and \emph{camel-artifact} for integrating the former agents with external entities modeled as environmental artifacts. 

\subsection{Apache Camel}\label{sec:Camel}

Apache Camel is a lightweight Java-based framework message routing and mediation engine~\cite{Ibsen:2010:CA:1965487}. Camel achieves high-performance processes handling multiple messages concurrently, and provides functions such as routing, exception handling, and testing. It uses structured messages and queues based on Enterprise Integration Patterns (EIP)~\cite{Hohpe:2003:EIP:940308}, preserving loose coupling among the resources. Camel works as a middleware that can be incorporated into an application through the use of \emph{components}. Communication among Camel components is defined in so-called \emph{routes}, which set and manage how messages will be exchanged, possibly following sets of rules and using data manipulation.

\begin{figure}[ht]
\centering
\includegraphics[width=0.8\textwidth]{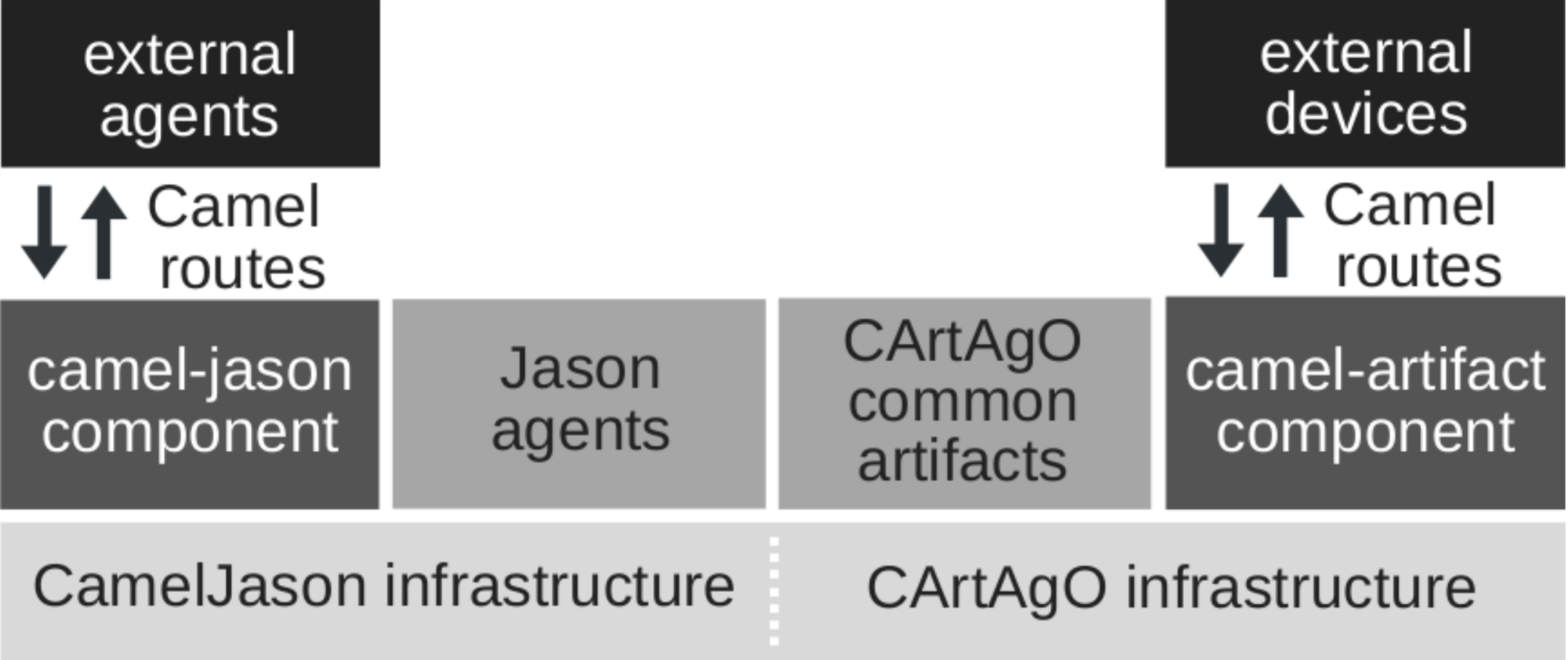}
\caption{MAS architecture using \emph{camel-jason} and \emph{camel-artifact} components.}
\label{fig:architecture}
\end{figure}

Routes define a single endpoint for each entity with an unique address. The defined endpoint may receive data, through a \emph{producer} and send data through a \emph{consumer}. Consumers are entities that admit data in specific formats and encapsulates it in a camel exchange object, an item that any other producer understands and is able to decode. Producers are entities that receive encapsulated data from the consumer decoding it in its entity's message structure.

In our implemented components, Camel is being embedded in two slightly different manners regarding the models of integration, A-A and A-E, as shown in Figure~\ref{fig:architecture}. In the case of A-A, it works as a communication infrastructure that is used when the recipient is not found locally. In the case of A-E, the external device is usually modelled reflecting real operations and signals that it generates, typically having their individuals routes. In both cases, the components are able to define tuned integration, covering a range of end-points features. Notice that, the complexity of each supported protocol is processed in a Camel component, which works as a bridge to Camel routes. There are more than two hundred components available on Camel's website\footnote{Supported Camel components are listed in http://camel.apache.org/components.html} and many others on the community's repositories.

\subsection{\emph{camel-jason} component}

The \emph{camel-jason} component enables agents to communicate with external entities through ACL, whilst fulfilling the need of understanding those entities as agents when modeling the MAS. In our proposal, the external entity has a kind of virtual counterpart inside the MAS, a \emph{dummy agent}. This counterpart is seen by the agents as an ordinary agent of the system. Doing so, agents can directly communicate with external entities assuming that they are other agents (as receivers and senders of ACL messages).  

\begin{figure}[ht]
\centering
\includegraphics[width=1\textwidth]{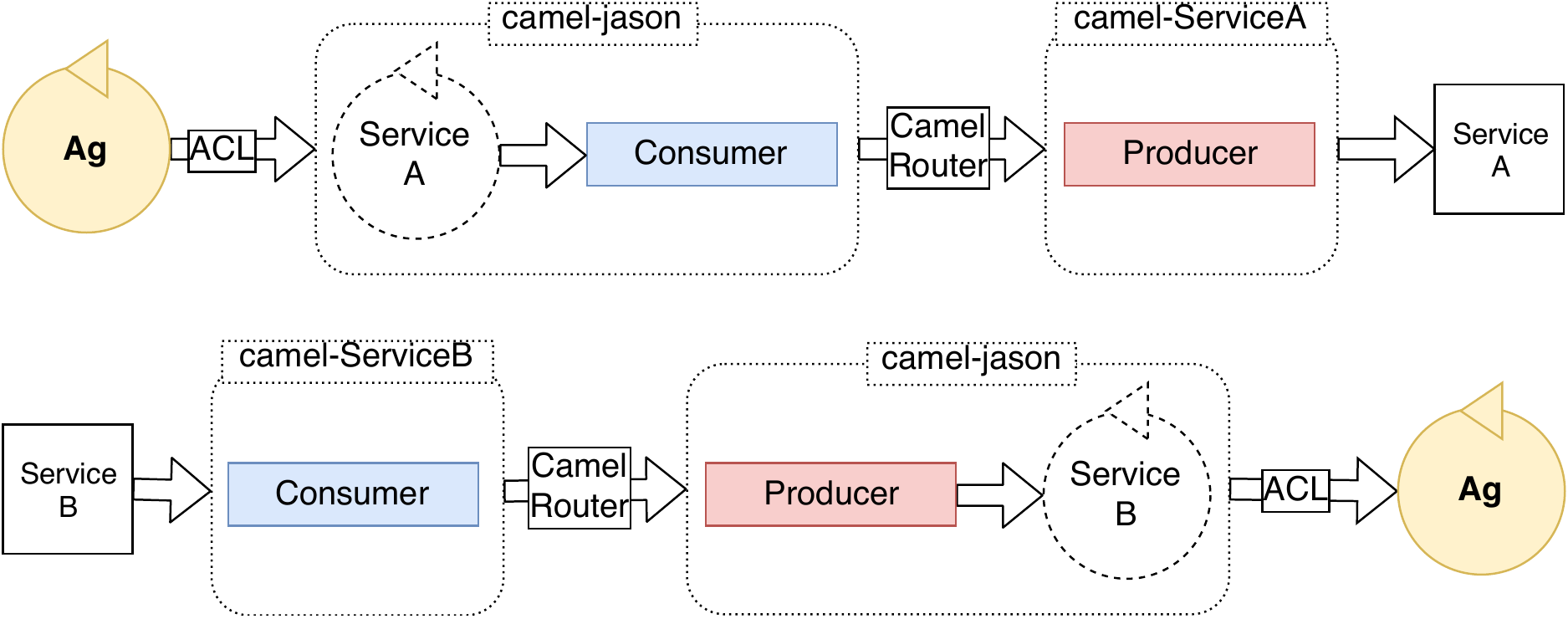}
\caption{Communication flow using \emph{camel-jason} component.}
\label{fig:cameljason}
\end{figure}

\emph{Camel-jason} component provides a communication flow that is illustrated in Figure~\ref{fig:cameljason} where an agent interact with a service A (an external entity). Since the agent sees the service as another agent, it uses ACL for the communication. When the service wants to contact the agent, the camel-jason component translates the message into ACL and the agent receives it as if it comes from an agent.

On the top communication flow showed in Figure~\ref{fig:cameljason}, we have an agent sending a message to an external autonomous entity: (i) the agent send an ACL message addressed to a dummy agent, which is created by \emph{camel-jason} component, referring to \emph{Service A}; (ii) the message is consumed by \emph{camel-jason} component consumer; (iii) the message is exchanged to the other side of the route, possibly being transformed; and (iv) the message is processed by \emph{Service A} component producer which prepares a service A compliance message, which will be sent to some network address to be effectively consumed by the \emph{Service A}.

In the other way around, on the bottom of Figure~\ref{fig:cameljason}, we have: (i) \emph{Service B} sends some data through the network reaching \emph{Service B} component consumer by its network address; (ii) the message is exchanged through Camel route, possibly being transformed; (iii) the message is processed by \emph{camel-jason} component producer which generates an ACL message; and (iv) the receiver agent effectively consumes the ACL message.

The component uses a simplistic method to define the communication routes, in which for many cases no actual programming is required, only XML definitions. The user should know how to fill camel endpoint parameters according to the compatible endpoint of the application. In cases data transformation is required, camel brings some tools for simple transformation as well as complex ones, using embedded programming codes if needed.

\subsection{\emph{camel-artifact} component}

In order to sustain the A-E model, the CArtAgO infrastructure is used, and the \emph{camel-artifact} component was developed. This component allows agents to perceive and act upon artifacts that represent external entities inside the MAS.

Notwithstanding, \emph{camel-artifact} also allows the definition of communication routes between CArtAgO artifacts and external entities. Routes for the \emph{camel-artifact} component are implemented using the Java language. The user should be aware of regular camel routes and how to define endpoints and their respective parameters.

\begin{figure}[ht]
\centering
\includegraphics[width=1\textwidth]{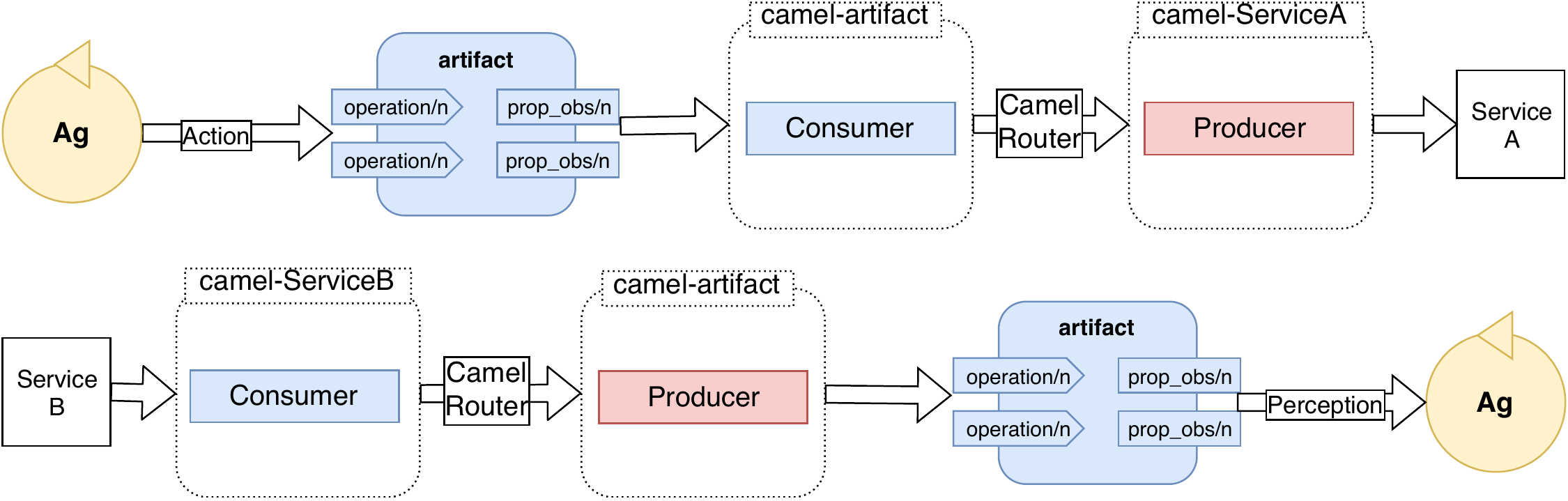}
\caption{Communication flow using \emph{camel-artifact} component.}
\label{fig:camelartifact}
\end{figure}

On the top communication flow showed in Figure~\ref{fig:camelartifact}, we have an artifact integrated with some service A, an external non-autonomous entity. The interaction between them is as follows: (i) the artifact sends a message to to \emph{Service A}, through specific methods provided by \emph{camel-artifact}; (ii) the message, in form of an operation request, is consumed by \emph{camel-artifact} component consumer which generates a camel standardised message to be sent to the external entity; (iii) the message is exchanged to the other side of the route, possibly being transformed; and (iv) the message is processed by \emph{Service A} compatible component producer which prepares a compliance final message, with the proper format and structure, which will be sent to some network address to be effectively consumed by the \emph{Service A}.

In the other way around, on the bottom of Figure~\ref{fig:camelartifact}, we have: (i) \emph{Service B} sends some data through the network reaching \emph{Service B} component consumer by its network address; (ii) the message is exchanged through Camel route, possibly being transformed; (iii) the message is processed by \emph{camel-artifact} producer which generates an artifact operation request; and (iv) the recipient artifact effectively consumes the operation request executing the referred method.

\section{Illustrative application}

For a better understanding of how the \emph{camel-jason} and \emph{camel-artifact} components can be used, we will resume the example of Industry 4.0, presented in Figure~\ref{fig:exampleFig1}, and will build an implementation of this system.

\begin{figure}[ht]
\centering
\includegraphics[width=1\textwidth]{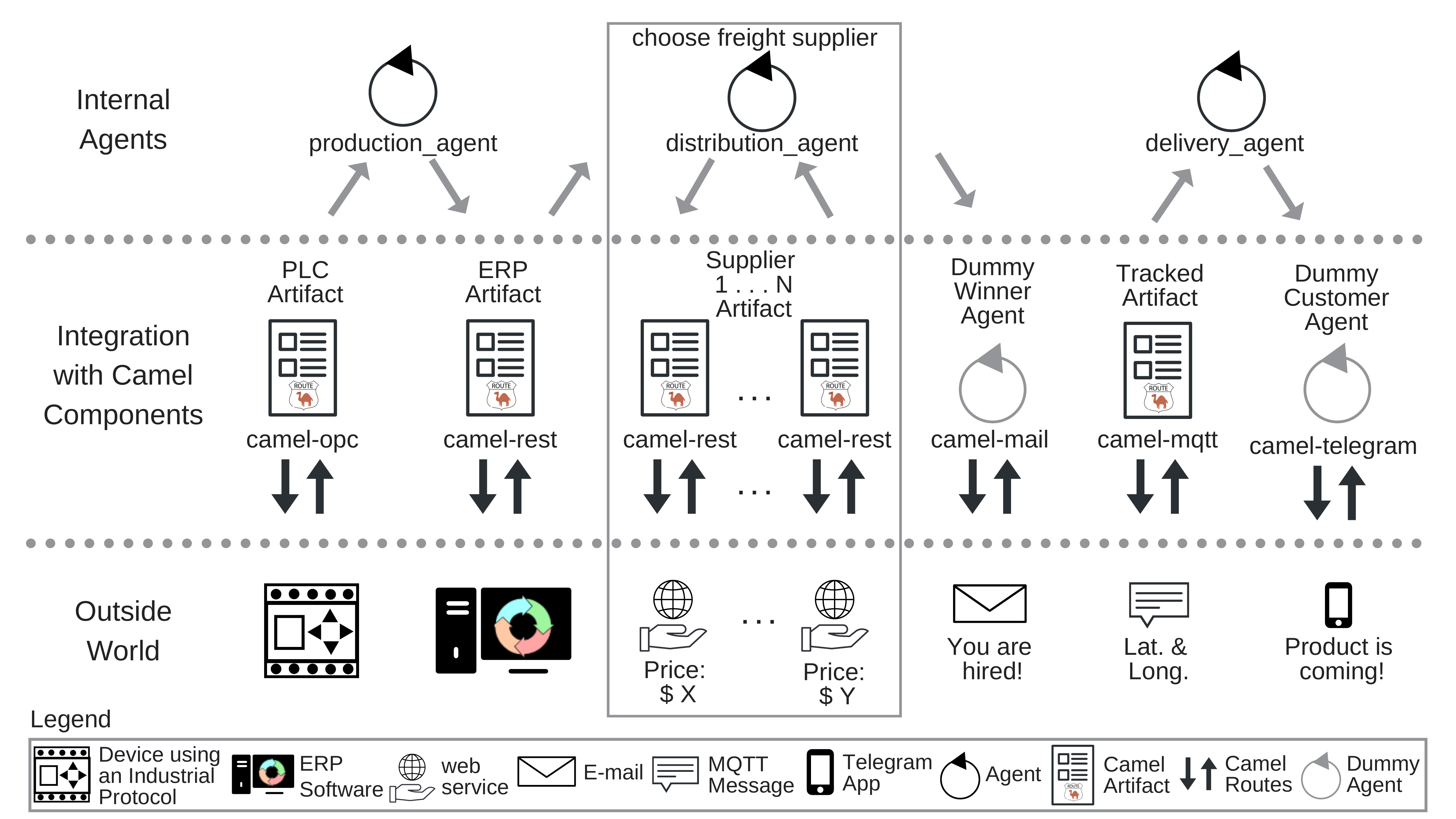}
\caption{An industrial process illustrating the integration of an MAS with a manufacturing device, updating an enterprise management software, choosing a supplier for delivering a tracked product to the customer.}
\label{fig:exampleFig2}
\end{figure}

The Figure~\ref{fig:exampleFig2} shows the MAS fully designed, with agents, external entities and the camel components used to implement the integration. These components are represented in the middle layer as artifacts and dummy agents. This hypothetical scenario implements an MAS to integrate the production and distribution stages of a product. The whole course can be divided in five stages: (i) a Programmable Logic Controller (PLC) finishes the product manufacturing, (ii) the information about the product is uploaded to an Enterprise Resource Planning (ERP) software, (iii) a research starts in order to contract the best freight company, (iv) the hired company starts transporting the product, providing its tracking information, and (v) warns the client via chat when it is near the final destination. The MAS is designed to unify those stages and to be responsible for managing each process. Moreover, Camel components are used as middleware between the MAS and external entities to integrate them.

One common question when designing the MAS is how many agents should be used. This is not mandatory, but a natural thought is to divide the process into sections and designate a single agent to be responsible for each part. In this case, we will consider that PLC and ERP stages represent the production part of the process, so one agent, named \emph{production\_agent}, will be responsible for managing these processes. Next, there is the hiring stage, that comprehends searching and hiring the best delivery company, for which the \emph{distribution\_agent} will be designated. The final stage could be thought as the delivery process, where the last agent, named \emph{delivery\_agent}, will be responsible for consulting the tracking information and sending the message to the costumer.        

Another part of the designing process is the identification of which model, agent or artifact, is more suitable to represent each external entity. A common way to decide is observing its nature, i.e. autonomous or non-autonomous. Following this idea, it could be decided that the PLC and ERP software would be modeled as artifacts, since they are non-autonomous entities; and the customer as an agent, an autonomous entity. Another possibility to decide is looking at which type of communication the agents will perform with each external entity, i.e. via message exchanging or perception-action. For example, the action of hiring the delivery company, informing the company about some new contract via email, seems natural to be modeled as an A-A interaction. For this situation, the best suitable performative for the message is \textit{tell}, since the agent is telling the supplier about a new hired delivery. On the other hand, when the \emph{distribution\_agent} is searching for the best delivery company, consulting their prices and conditions, it seems more suitable for this information to be perceived, like people do in a websearch. The same idea can be used when the \emph{delivery\_agent} tracks the position of the product, the information again is perceived by the agent, like looking at a screen. If message exchange was used in this case the agent would be flooded with unnecessary messages.

In fact, both interpretations, i.e., perception-action vs message exchanging and autonomous vs non-autonomous, could point to the same conclusion. This statement can be tested in the integration between the \emph{delivery\_agent} and the customer. The chat is done by message exchanging and it is performed by autonomous entities. Therefore, both interpretation reinforces the idea that the costumer should be modeled as an agent.

With the external entities modeled as MAS elements we could use \emph{camel-artifact} and \emph{camel-jason} to integrate the internal agents with the artifacts and agents, respectively. It is worth commenting that the external entities could be easily exchanged, for instance making the auction via email instead of using a web service. The interaction via email suggests modelling the participants as agents and use the \emph{camel-jason} component for the integration.

Now the camel routes can be developed, depending on which type of technology the external entities use. As explained before, Camel have more than two hundred endpoints available. In this example, OPC-DA, Rest, email, MQTT and Telegram end-points are being used to create the routes.

The code, in XML, for the route from the \emph{delivery\_agent} to Telegram can be seen in Listing~\ref{lst:CamelJasonRoute}. The \code{from} tag signalises the consumer part of the route, and \code{to} signalises the producer. In this case, the consumer address is the name of the dummy agent to which the message had been sent (in this example, \code{customer}), and the producer address is the authorisation token followed by the \code{chatId} option.

\begin{lstlisting}[caption={Example of \emph{camel-jason} route definition}, captionpos=b, label=lst:CamelJasonRoute, frame=none, framexleftmargin=0pt, numbers=left, numbersep=5pt, numberstyle=\tiny, language=XML, showspaces=false, showtabs=false, breaklines=true, showstringspaces=false, breakatwhitespace=true, basicstyle=\footnotesize\ttfamily,
morekeywords={route, from, to, uri, setHeader}]
<route>
  <from uri="jason:DummyCustomerAgent"/>
  <to uri="telegram:bots/sometoken?chatId=-364531"/>
</route>   
\end{lstlisting}

In this example, the \emph{delivery\_agent} also uses the \emph{camel-artifact} in order to obtain the delivery's position from a MQTT server. The route, in Java, is shown in Listing~\ref{lst:CamelArtifactRoute}. When the position is published on the topic of interest (\code{latLong}) the route redirects it to the artifact by specifying its name (\code{TrackedArtifact}) and the operation (\code{giveDistance}) as headers. Here, we are assuming that the calculations will be done by the artifact, but they could be done in the route through a transformation, before sending to the artifact.

\begin{lstlisting}[caption={Example of \emph{camel-artifact} route definition}, captionpos=b, label=lst:CamelArtifactRoute, frame=none, framexleftmargin=0pt, numbers=left, numbersep=5pt, numberstyle=\tiny, language=Java, showspaces=false, showtabs=false, breaklines=true, showstringspaces=false, breakatwhitespace=true, basicstyle=\footnotesize\ttfamily,
morekeywords={route, from, to, setHeader}]
from( "mqtt : foo? host=tcp://broker & subscribeTopicName=latLong" )
  .setHeader( "ArtifactName" , constant ( "TrackedArtifact" ) )
  .setHeader( "OperationName" , constant ( "giveDistance" ) )
.to( "artifact : cartago" );
\end{lstlisting}

\section{Related research}

In this section, we went over works that have addressed agent technology in an integrating context. Maturana and Norrie~\cite{Maturana1996} have proposed a mediation and coordination tool for MAS. They have used mediator agents as manufacturing coordinators. Following similar idea, Olaru et al.~\cite{Olaru2013} have developed an agent-based middleware, which creates a sub-layer of application layer that allows agents to mediate context-aware exchange of information among entities. We think an autonomous entity as middleware may increase complexity and compromise performance. Instead of creating some kind of hierarchy, our approach gives connectivity power to MAS entities.

Leading industrial suppliers are also providing solutions using agents such as the Agent Development Environment (ADE), designed by Rockwell Automation~\cite{TICHY2012846}. It provides connectivity with common shop floor devices and supports the development of agents. The limitation we have seen regards especially connectivity with all sorts of entities (e.g. IoT sensors and mobile devices, ERP and other software etc), which in our case is provided by Camel.

Other research address the combination of MAS and Service-Oriented Architecture (SOA). One way to achieve this merge is based on the creation of a proxy function to provide interoperability between MAS and SOA, as found in \cite{Nguyen2005, shafiq2005agentweb, greenwood2004,Faycal2010}. Another way is by implementing services as agents as we found in \cite{Mendes2009,tapia2009fusion,carrascosa2009thomas, argente2011thomas}. The approaches using SOA are more mature to be applied in practice. The ones that \emph{agentified} the services have also the advantage to use MAS background, i.e., using ACL messages they are able to use interaction protocols. In these studies integration is usually done through specific APIs and they lack differentiation over autonomous and non-autonoumous entities, and interoperability with heterogeneous entities, both aspects increase development complexity.

Vrba et al. \cite{Vrba2014} propose a gateway for wrapping an MAS as a service to be used as a loosely coupled software component into the Enterprise Service Bus (ESB). This gateway transforms agent messages to ESB messages and vice versa, enabling communication between agents and ESB services. This solution is closely to ours, only lacking support to the A-E approach, where interaction is based on agents perceiving and acting upon artifacts in the environment. The indiscriminately \emph{agentification} may increase complexity and affect performance.

Cranefield and Ranathunga~\cite{Cranefield2013} developed a \emph{camel-agent} component for \emph{Jason} agents. It is very similar to our developed \emph{camel-jason} component. Essentially, the difference is that we are embedding Apache Camel since our component works as an infrastructure, being transparent to the agents. In their work, Jason was actually embedded in an Apache Camel project where agents were smoothly placed in containers.

\section{Conclusion}

In this paper, we introduced two Camel components aiming the integration of MAS with external entities: \emph{camel-jason} and \emph{camel-artifact}. The former integrates agents with external entities modelled as agents. The latter integrates agents and external entities modelled as artifacts. The decision of which component to adopt for each entity depends on the characteristics of the external entity and the MAS developer \emph{can} choose the most suitable component. For instance, he/she is not obliged to ``agentify'' every external entity, even those that do not have agent properties.

The two components introduced in this paper, along with the communication infrastructure provided by Camel and its existing components, makes the integration between MAS and different entities simpler. Issues related to interoperability, routing, and data transformations are partially solved in the camel routes. Another advantage of using such components is that the agent program does not need to deal with integration issues. Agents continue to interact only with another agents and artifacts.

Finally, this is an ongoing work. In a future step we  intend to compare our approach with related works and to evaluate other aspects of using the developed Camel components to integrate MAS and external entities, such as the impact on the performance, security, openness, scalability, among others.

\bibliographystyle{sbc}
\bibliography{sbc-template}

\end{document}